\documentclass{jac}

\usepackage{amssymb}
\usepackage{amsmath}
\usepackage{ amsfonts} 

\usepackage{stmaryrd}
\usepackage{mathrsfs}
\usepackage{latexsym}
\usepackage{hhline}

 \newcommand{\comment}[1]{{}}

\begin{document}

\title [Tiling-recognizable Languages]{Tiling-recognizable Two-dimensional Languages: from Non-determinism to Determinism\\ through Unambiguity}

\author
{D. Giammarresi}{Dora Giammarresi}%
\address
{  Dipartimento di Matematica,
 Universit\`a  di Roma ``Tor Vergata''\\
  via della Ricerca Scientifica\\ 00133 Roma, Italy}
 \email{giammarr@mat.uniroma2.it}

\thanks{This work was partially supported
by ESF Project {\it ``AutoMathA''} (2005-2010).  }


\hyphenation{two-di-men-sio-nal Two-di-men-sio-nal}
\hyphenation{de-ter-mi-ni-stic}
\hyphenation{non-de-ter-mi-ni-stic}
\hyphenation{lan-gua-ge}
\hyphenation{ge-ne-ra-li-za-tion}
\hyphenation{one-di-men-sio-nal}
\hyphenation{qua-dru-ple}
\hyphenation{con-straints}
\hyphenation{clo-su-re}
\hyphenation{ro-ta-tion}
\hyphenation{dia-go-nal}
\hyphenation{unam-bi-gui-ty}
\hyphenation{si-mi-lar}
\hyphenation{ar-gu-ment}

\begin{abstract}\noindent
Tiling recognizable two-dimensional languages, also known as REC, generalize
recognizable string languages to two dimensions and share with
them several theoretical properties. Nevertheless REC is  not closed under complementation and the membership problem is NP-complete. This implies that
this family REC is intrinsically non-deterministic.  The  natural and immediate definition of unambiguity  corresponds to a family UREC of languages that is strictly contained in REC. On the other hand this definition of unambiguity  leads to an undecidability result and therefore it cannot correspond to any deterministic notion. We introduce the notion of   line-unambiguous tiling recognizable languages  and prove that it corresponds  or somehow naturally introduces   different notions of  determinism  that define a  hierarchy inside REC.
\end{abstract}

 \maketitle


A \emph{picture} (or two-dimensional string) is a two-dimensional
arrays of symbols from a finite alphabet. A set of pictures is called  \emph{two-dimensional language}.  Basic notations and operations can be extended from string to pictures. The size of a picture $p$ is a pair $(m,n)$ corresponding to the number of its rows and columns, respectively. Moreover there can be defined an operation of column-concatenation between pictures with the same number of rows and of row-concatenation between pictures with the same number of columns. By iteration, there can be also defined the corresponding row- and column- star operations.

The first generalization of
finite-state automata to two dimensions can be attributed to M.~Blum and
C.~Hewitt who in 1967  introduced the notion of a four-way
automaton moving on a two-dimensional tape as the natural extension of
a one-dimensional two-way finite automaton (see~\cite{BH67}). They also proved that the deterministic version corresponds to a language class smaller than the corresponding one defined by the non-deterministic model. Four-way automata   was not a successful model since the corresponding language class does not  satisfies important properties as closure under  concatenation and star operations. Since then, many approaches have been presented
in the literature in order to find the "right way" to generalize in 2D  what
regular languages are in one dimension: finite automata,
grammars, logics and regular expressions (see for example \cite{BG04,GR91,M98,IN77,S87}). Here we focus on the family \emph{REC} of {\em tiling recognizable picture languages}
(see \cite{GR91,GR-book}  that have been widely investigated and that it is considered as a valid candidate to represent a counter part to 2D of regular string languages.

 The definition of REC takes as
starting point a characterization of recognizable string languages
in
 terms of local languages and projections (cf. \cite{E74}). A picture language $L$ is \emph{local} if it is defined by a finite set of $2\times 2$ pictures, called \emph{tiles} that represent all allowed sub-pictures of size $(2,2)$ for   pictures in $L$. A pair composed by a local language over an alphabet $\Gamma$ and an alphabetic projection $\pi:\Gamma \longrightarrow \Sigma$ is called \emph{tiling system}. A picture language $L$ over an alphabet $\Sigma$ is recognized by a tiling system (given by a local language  $L'$ and $\pi$)   if each picture $p\in L$ can be obtained as projection of a picture $p'\in L'$ ( i.e. $p=\pi(p')$). A picture language is tiling recognizable if it is recognized by a tiling system. REC is the family of tiling recognizable picture languages. We point that
languages of infinite picture  ($\omega$-pictures)
were also studied in the setting of tiling systems in \cite{AltenberndTW02,Finkel04,Finkel09}.

 It can be verified that REC is closed under union and intersection, rotation and mirror and under column- and row- concatenation and star operations.
Moreover,
the definition of REC in terms of tiling systems turns out to be
very robust:
in~\cite{GR-book,GRST94} it is shown that the family REC has a characterization in terms of
logical formulas (a generalization of
B\"uchi's theorem for strings to 2D). In~\cite{IT92}, it is proved that REC has a counterpart as machine model in the {\em
two-dimen\-sional on-line tessellation acceptor (OTA)}
introduced by K.~Inoue and A.~Nakamura
in~\cite{IN77}. Other models of automata for REC are proposed in \cite{AGM07-tila,BG04,LP10}.
Tiling systems can be also simulated by domino systems \cite{IT92} and Wang tiles \cite{DPV97} and grammars \cite{CRP05}. Further we remark that when pictures degenerate in strings (i.e. when considering only one-row pictures) recognizability by tiling systems corresponds exactly to recognizability by finite state string automata.

 A crucial difference with the one-dimensional case
lies in the fact
that the definition of recognizability by tiling systems is
intrinsically non-deterministic.  Deterministic machine models to
recognize  two-dimensional languages have been considered in the
literature: they always accept classes of languages smaller than
the corresponding non-deterministic ones (see  for example,
\cite{BH67,IN77,PST94}). This seems to be unavoidable when jumping
from one to two dimensions. Further REC family is not closed under
complementation and therefore the definition of any constraint to force
determinism in  tiling systems should necessary result in a class
smaller than REC.
Strictly connected with this problems are the complexity results on the recognition problem in REC. Let $L$ be a language in REC defined by a tiling system composed by a local picture language $L'$ and a projection $\pi$.  To recognize that a given picture $p$ of $m$ rows and $n$ columns belongs to $L$, one has  to "rewrite"  symbols in all positions in $p$ to get a local picture $p'$ that belongs to $L'$ and such that $\pi(p')=p$. This can be done by scanning all positions of $p$ in some order. The non-determinism implies that, once reached  a given position  one may eventually backtrack on all positions already visited, that is on $O(mn)$ steps.
Moreover in \cite{LMN98} it is proved that the
recognition problem for REC languages is NP-complete.

In formal language theory, an intermediate notion between
determinism and non-determinism is the notion of  unambiguity. In
an unambiguous model, we require that each accepted object admits
only one successful computation. Both determinism and unambiguity
correspond to the existence of a  \textit{unique} process of
computation, but while determinism is a ''local'' notion,
unambiguity is a fully ''global'' one. \emph{Unambiguous tiling recognizable}
two-dimensional languages have been introduced in \cite{GR91}, and
their family is referred to as UREC. Informally, a picture
language belongs to UREC if it admits an unambiguous tiling
system, that is if every picture has a unique pre-image in its
corresponding local language.
In  \cite{AGMR}, the proper inclusion of UREC in REC is
proved but  it is also proved that
it is undecidable  whether a given tiling system is unambiguous. From a computational side, there are not known  algorithms to recognize pictures that exploit the properties of UREC. This implies that,   at each step of the recognition computation, it can be necessary to backtrack on all already visited positions.

A relevant goal is then to  find subclasses for REC that inherit important properties but also allow  feasible computations. Moreover an interesting result would be  proving that, as for regular string languages,  notions of some kind of unambiguity and determinism coincide.

Remark that another difference between unambiguity and determinism is that determinism is always  related to a scanning strategy to read the input. In the string case the scanning is implicitly assumed to be left-to right and in fact deterministic automata are defined related to this direction. Moreover since deterministic, non-ambiguous and non-deterministic models are all equivalent there is no need to consider determinism from right-to-left (referred to as co-determinism). Nevertheless it is worthy to remark that not all regular string languages admits automata that are both deterministic and co-deterministic. In the two-dimensional case we have to consider all the  scanning directions from left, right, top and bottom sides.

By exploiting the different possibilities of scanning for a two-dimensional array in \cite{AGM07,AGM09} there are introduced different notions of unambiguity we call here \emph{line-unambiguity} where a line can be either a column or a row or a diagonal. We consider tiling systems for which the  computations to recognize a given picture can have at each position a backtracking on at most $m+n$ steps.
Such definitions lie between those of unambiguity and determinism (as long as we consider that a deterministic computation has zero backtracking steps at each position) while they all coincide with determinism when pictures degenerate in strings.

The informal definitions are very simple and natural.
A tiling system is \emph{column-unambiguous}  if,
when
 used to recognize a picture by reading it
along a left-to-right or right-to left direction, once computed a local column, there is only one possible next local column. As consequence in a computation by
 a column-unambiguous tiling system  to recognize a picture with $m$ rows, the
backtracking at each
step is at most of $m$ steps.
Similarly there are defined  \emph{row-unambiguous} and  \emph{diagonal-unambiguous} tiling systems corresponding to computations that proceed  by rows or by diagonals, respectively. The corresponding families of languages are denoted by {\em Col-UREC}, {\em Row-UREC} and {\em Diag-UREC}.
In  \cite{AGM07,AGM09} there are proved
necessary conditions for a language to be in Col-UREC and in Row-UREC. Using such
conditions  one can show that  families Col-UREC and Row-UREC are strictly contained in UREC.
In a different set-up it is also shown that Diag-UREC is strictly included both in Col-UREC and Row-UREC. Moreover all those properties are decidable.

Very interestingly we can  prove that diagonal-unambiguous tiling systems are  equivalent to some deterministic tiling systems where the uniqueness of computation is guaranteed by certain  conditions on the set of local tiles: the corresponding language family is denoted by DREC (\cite{AGM07}). Similar results hold for classes Col-UREC and Row-UREC whose union turns to be equivalent to another "deterministic" class named Snake-DREC \cite{LP09}.
All those classes are closed under complementation \cite{AGM09,LP09}. As result,  when we consider this  line unambiguity we can prove equivalence with  deterministic models and therefore  we  guarantee  a recognition algorithm linear in the size (i.e. number of rows times number of columns) of the input.


\begin{thebibliography}{10}
\bibitem{AltenberndTW02}
J.-H. Altenbernd, W.~Thomas, and S.~W{\"o}hrle.
\newblock Tiling systems over infinite pictures and their acceptance
  conditions.
\newblock In {\em Developments in Language Theory 2002}, volume 2450 of {\em
  Lecture Notes in Computer Science}, pages 297--306. Springer, 2003.



\bibitem{AGM09} M. Anselmo, D. Giammarresi, M. Madonia.
M.~Anselmo, D.~Giammarresi, and M.~Madonia.
\newblock Deterministic and unambiguous families within recognizable
  two-dimensional languages.
\newblock {\em Fundamenta Informaticae}, 98(2-3):143--166, 2010.

\bibitem{AGM07} M. Anselmo, D. Giammarresi, M. Madonia.
\newblock From determinism to non-determinism in recognizable two-dimensional languages.
\newblock In {\em Procs. DLT 07}, T. Harju, J. Karhumaki  and A. Lepisto (Eds.), LNCS 4588,
Springer-Verlag, Berlin 2007.

\bibitem{AGM07-tila} M. Anselmo, D. Giammarresi, M. Madonia.
\newblock  A computational model for tiling recognizable two-dimensional languages.
\newblock {\em Theoretical Computer Science}, Vol. 410-37, 3520--3529 Elsevier 2009.

\bibitem{AGMR} M. Anselmo, D. Giammarresi,
M. Madonia, A. Restivo. \newblock Unambiguous Recognizable
Two-dimensional Languages. {\em RAIRO: Theoretical Informatics and
Applications}, Vol. 40, 2, pp. 227-294, EDP Sciences 2006.

\bibitem{AM08} M. Anselmo,
M. Madonia. \newblock Deterministic and unambiguous two-dimensional languages over one-letter alphabet.
\newblock {\em Theoretical Computer Science}, Vol. 410-16, 1477--1485 Elsevier 2009.



\bibitem{BH67}
M. Blum, C. Hewitt.
\newblock  Automata on a two-dimensional tape.
\newblock {\em IEEE Symposium on Switching and Automata Theory}, pages
155--160, 1967.




\bibitem{BG04} S. Bozapalidis, A. Grammatikopoulou,
\newblock Recognizable picture series,
\newblock {\em Journal of Automata, Languages and Combinatorics}, special vol. on
{\em Weighted Automata, 2004}.

\bibitem{CRP05} S. Crespi Reghizzi and M. Pradella.
\newblock Tile rewriting grammars and picture languages.
   \newblock {\em
Theoretical Computer Science}, vol 340, n.2, pp. 257-272, Elsevier
2005.


\bibitem{DPV97}
De Prophetis, L., Varricchio, S.: Recognizability of rectangular
pictures by wang systems. Journal of Automata, Languages,
Combinatorics. {\bf 2} (1997) 269-288

\bibitem{E74}
S. Eilenberg.
\newblock {\em Automata, Languages and Machines}.
\newblock Vol. A, Academic Press, 1974.

\bibitem{Finkel04}
O.~Finkel.
\newblock On recognizable languages of infinite pictures.
\newblock {\em Int. J. Found. Comput. Sci.}, 15(6):823--840, 2004.

\bibitem{Finkel09}
O.~Finkel.
\newblock Highly undecidable problems about recognizability by tiling systems.
\newblock {\em Fundam. Inform.}, 91(2):305--323, 2009.

\bibitem{GR91}
D.~Giammarresi, A.~Restivo.
\newblock Recognizable picture languages.
\newblock \textit{Int. Journal Pattern Recognition and Artificial
Intelligence.}
\newblock Vol. 6, No. 2\& 3, pages 241 --256, 1992.


\bibitem{GR-book}
D.~Giammarresi, A.~Restivo.
\newblock Two-dimensional languages.
\newblock {\em Handbook of Formal Languages}, G.Rozenberg,
\textit{et al.} Eds, Vol. III, pag. 215--268. Springer Verlag,
1997.

\bibitem{GR07} D.~Giammarresi, A.~Restivo. Matrix-based
complexity functions and recognizable picture languages.
\newblock In \emph{Logic and Automata:
History and Perspectives.}  E. Grader, J.Flum, T. Wilke Eds. ,pag
315-337. Texts in Logic and Games 2.  Amsterdam University Press,
2007.

\bibitem{GRST94}
D. Giammarresi,  A. Restivo, S. Seibert, W. Thomas.
\newblock Monadic second order logic over pictures and
recognizability by tiling systems.
\newblock {\em Information and Computation}, Vol 125, 1, pag 32--45,
1996.


\bibitem{IN77}
K.~Inoue, A.~Nakamura.
\newblock Some properties of two-dimensional on-line tessellation
acceptors.
\newblock {\em Information Sciences}, Vol. 13, pages 95--121, 1977.

%
\bibitem{IT92}
K.~Inoue, I.~Takanami.
\newblock A characterization of recognizable picture languages.
\newblock In {\em Proc. Second International Colloquium on Parallel Image
Processing},  A.~Nakamura et al. (Eds.), LNCS 654,
Springer-Verlag, Berlin 1993.

\bibitem{LS94} M.~Latteux and D.~Simplot.
\newblock Recognizable Picture Languages and Domino Tiling. \emph{Theorethical Computer Science} 178(1-2): 275-283, 1997.

\bibitem{LMN98}   
K.~Lindgren, C.~Moore, M.~Nordahl.
\newblock Complexity of two-dimensional patterns.
\newblock {\em Journal of Statistical Physics}, 91 (5-6), pag. 909--951, 1998.
\bibitem{M97}
O. Matz.\newblock Regular expressions and Context-free Grammars
for picture languages. {\em Proc. STACS'97 }- LNCS 1200 pag.
283-294 - Springer Verlag 1997.

\bibitem{LP09}
 V. Lonati, M. Pradella.
\newblock Snake-Deterministic Tiling Systems.
\newblock In \emph{Proc. MFCS 2009},
LNCS, Vol. 5734,
549-560, Springer 2009.

\bibitem{LP10}
V. Lonati and M. Pradella.
\newblock Picture-recognizability with automata based on Wang tiles.
\newblock In \emph{Proc. SOFSEM
2010}, LNCS, vol. 5901,    576-–587. Springer, 2010.

\bibitem{M98}
O. Matz. \newblock  On piecewise testable, starfree, and
recognizable picture languages. In {\em Foundations of Software
Science and Computation Structures}, M. Nivat Ed., vol. 1378,
Springer, 1998.
%



\bibitem{PST94}
A.~Potthoff, S.~Seibert, W.~Thomas.
\newblock Nondeterminism versus determinism of finite automata
over directed acyclic graphs.
\newblock {\em Bull. Belgian Math. Soc.} 1, 285--298, 1994.

\bibitem{S87}
R.~Siromoney.
\newblock Advances in array languages.
\newblock In {\em Graph-Grammars and Their Applications to Computer
Science}, Ehrig et al. (Eds.), pages 549--563.
\newblock Lecture Notes in Computer Science 291,
Springer-Verlag, Berlin, 1987.

\end{thebibliography}
\end{document}